\documentstyle[sprocl]{article}

\input{psfig}

\def\drop#1{}

\def\equ#1{~Eq.~(\ref{#1})}

\newcommand{\vev}[1]{\left\langle #1 \right\rangle}

\newcommand{\sla}{\raise.15ex\hbox{$/$}\kern -.8em} 
 
\newcommand{\half}{\frac 1 2}

\newcommand{\cb}{{\bar c}}

\renewcommand{\L}{\Lambda}
\newcommand{\LQCD}{\L_{\overline{MS}}}

\newcommand{\m}{\mu}\newcommand{\n}{\nu}
\newcommand{\mn}{{\mu\nu}}
\newcommand{\eps}{\varepsilon}
\newcommand{\bra}{\langle}\newcommand{\ket}{\rangle}

\newcommand{\be}[1]{\begin{equation}\label{#1}} 
\newcommand{\ee}{\end{equation}}
\newcommand{\ba}[1]{\begin{eqnarray}\label{#1}} 
\newcommand{\ea}{\end{eqnarray}}

\begin{document}

\title{SU(2) GAUGE THEORY IN COVARIANT (MAXIMAL) ABELIAN GAUGES}

\author{M. SCHADEN}

\address{New York University, Physics Department,\\
4 Washington Place, New York, NY 10003-6621\\
E-mail: ms68@scires.nyu.edu}

\maketitle\abstracts{The local covariant continuum action of an SU(2)
gauge theory in covariant Abelian gauges is investigated. It  
describes the critical limit of an Abelian Lattice Gauge Theory (LGT) 
with an equivariant BRST-symmetry.  This Abelian LGT has previously
been proven to be physically equivalent to the SU(2)-LGT. 
Renormalizability requires a quartic ghost interaction in these 
non-linear gauges (also in maximal Abelian gauge). Arguments that 
a certain global SL(2,R) symmetry is dynamically broken by ghost-antighost 
condensation in a BCS-like mechanism are presented. The scenario can
be viewed as a dynamical Higgs mechanism in the adjoint that gives 
massive off-diagonal gluons and a BRST quartet of Goldstone bosons
that decouples from physical observables. The gap parameter is related 
to the expectation value of the trace anomaly and the consistency of this 
scenario with the Operator Product Expansion is discussed.}

\section{Introduction}

An $SU(2)$ Lattice Gauge Theory (LGT) on a finite lattice is invariant
under the compact group ${\cal G}$  
\be{group}
{\cal G}=\otimes_{\rm sites} SU(2)\ .
\ee
${\cal G}$ does not have a smooth continuum limit and it is not
entirely clear whether some of the effects observed on the lattice, such as
absolute confinement, are due to the compact nature of ${\cal G}$ and
are absent in a local Quantum Field Theory (QFT) with non-compact 
$SU(2)$ gauge invariance. The fact that compact (lattice) QED {\it
does} differ markedly from the continuum theory due to lattice
``artefacts'' makes this question all the more  
relevant. These ``artefacts'' are Abelian monopoles specific to lattice QED --
their existence is intimately related to
the compactness of the (Abelian) lattice gauge group and they have no
continuum analogues. In the
Abelian case, one can remove these lattice artefacts by imposing the 
constraint
\be{proj}
A_\mu=\bigtriangleup^{-1}\partial_\mu F_{\mu\nu}\ ,
\ee
where $P_\mn(x)=e^{i F_{\mu\nu}(x)}$ is the plaquette variable and
$U_\mu(x)=e^{i A_\mu(x)}$ is the $U(1)$ link variable. 
On the lattice,\equ{proj} is not just a gauge fixing condition, but
in addition eliminates the monopoles associated with
harmonic one forms. The continuum limit of this (projected) Abelian
LGT is free QED in Landau gauge. If\equ{proj} was {\it not} imposed,  
the gauge invariant transverse photon correlation function of 
lattice QED was found to differ markedly\cite{Fo95} from what
one expects in the continuum.

In view of this example, the question whether the critical limit of a
non-Abelian LGT can be described by ``QCD'' with a non-compact gauge
group is legitimate. It has been conjectured\cite{Pa99} that a
non-Abelian  LGT may not be asymptotically free and exhibit a
Kosterlitz-Thouless phase transition  at a {\it finite} value of the
coupling constant. 

In resolving the issue of the critical limit of a non-Abelian LGT it may 
be useful to have a physically equivalent local LGT with an
equivariant BRST-symmetry whose structure group has been reduced to
the maximal Abelian subgroup of the original LGT. The two LGT's in question 
are physically equivalent because the expectation
values of gauge-invariant operators (i.e. Wilson loops and their
linked generalizations) are the same. 
Although this equivalent Abelian LGT has only been
constructed for an $SU(2)$-LGT\cite{Sc99}, the method can be generalized to any
$SU(n)$-LGT. The lattice group ${\cal G}$ of an
$SU(2)$-LGT  is reduced to the maximal Abelian subgroup ${\cal H}$ 
by a local Topological Lattice Theory (TLT) that computes
the Euler number of the coset ${\cal G}/{\cal H}$,
\be{euler} 
\chi({\cal G}/{\cal H})=\otimes_{\rm sites}\chi(SU(2)/U(1)\sim
S_2)=2^{\rm \# sites}\ ,
\ee
on each orbit of a lattice configuration using Morse Theory. I should
stress that this construction of a TLT is conceptually quite different
from the usual Faddeev-Popov procedure and does not require the
uniqueness of the solution to a ``gauge condition'' -- the Euler
number of the manifold would in fact have to be~1 for this to be the
case. There are at least $2^{\rm \# sites}$ gauge equivalent Gribov copies
that contribute to $\chi({\cal G}/{\cal H})$ on any orbit of the
(finite) lattice.  The construction of the TLT is mathematically
 rigorous, because the coset manifold ${\cal G}/{\cal H}$ is compact and 
finite-dimensional on a finite lattice (albeit of rather large
dimension) and the orbit-space of the original LGT is connected. One
cannot reduce the full lattice group ${\cal G}$ in this manner because 
$\chi({\cal G})=0$. The best one can do is to reduce the lattice gauge
group to the smallest subgroup ${\cal H}$ for which the Euler number
of the coset manifold ${\cal G}/{\cal H}$
does not vanish. In the case of compact $SU(n)$ the smallest subgroup which
satisfies this requirement is the maximal Abelian one.    

Of interest here will be the continuum theory that describes the critical
limit of this ``partially gauge fixed'' LGT. 
The equivariant BRST-symmetry and $U(1)$-invariance together with
locality and power-counting renormalizability determine the continuum 
limit up to lattice artefacts associated
with the compactness of the  residual $U(1)$-structure group.  
Assuming these Abelian artefacts have been removed in a manner similar
to the one prescribed above, the critical limit of this LGT 
is unique because the BRST-invariance of the LGT is a global one.
The continuum model is then  described by the local action given below.    

Because physical correlation functions are the same\cite{Sc99}
in the reduced Abelian LGT and can be shown to satisfy reflection
positivity\cite{Se82} in the original $SU(2)$-LGT,  
the physical states of the ``partially 
gauge fixed'' Abelian LGT also have positive norm. By proving the
equivalence of the two LGT's for gauge-invariant correlation
functions one thus also verifies the unitarity of the
partially gauge-fixed Abelian LGT. The continuum theory describing the
critical limit of this LGT should therefore be
unitary as well. Note that this proof of the unitarity of the
continuum theory is valid non-perturbatively and not just to all
orders in perturbation theory. Instead of investigating the critical
limit of the original $SU(2)$-LGT, we thus will consider the critical limit of 
the (physically) equivalent LGT with an Abelian structure group
and an equivariant BRST-symmetry. 

\section{Continuum SU(2) Gauge Theory in Abelian Gauges}
Up to Abelian lattice artifacts mentioned above, the continuum theory
describing the critical limit of the ``partially 
gauge fixed'' Abelian LGT\cite{Sc99} is completely specified by the
global symmetries and power counting. It is described by the Lagrangian
\be{L}
{\cal L}={\cal L}_{\rm inv.}+{\cal L}_{\rm AG}+{\cal L}_{\rm aGF}\ .
\ee
Here ${\cal L}_{\rm inv.}$ is the usual $SU(2)$-invariant Lagrangian with the 
SU(2)-connection $\vec V_\m=(W_\m^1,W_\m^2,A_\m)$ written\footnote{Latin
indices take values in $\{1,2\}$ only, Einstein's summation
convention applies and $\eps^{12}=-\eps^{21}=1$, vanishing 
otherwise. All results are given in the $\overline{MS}$ renormalization  
scheme.} in terms of two (real) vector bosons $W$, and an 
Abelian  ``photon''  $A$, 
\be{Linv}
{\cal L}_{\rm inv.}={\cal L}_{\rm matter}+ {\frac 1 4} (G_\mn G_\mn+ G_\mn^a
G_\mn^a)\ ,
\ee 
with
\ba{G}
G_\mn&=&\partial_\m A_\n-\partial_\n A_\m -g\eps^{ab} W_\m^a W_\n^b\nonumber\\
G_\mn^a&=&D^{ab}_\m W_\n^b- D^{ab}_\n W_\m^b =\partial_\m W_\n^a
-\partial_\n W_\m^a+g\eps^{ab}(A_\m W_\n^b-A_\n W_\m^b)\ .
\ea
${\cal L}_{\rm AG}$ reduces the invariance to the maximal
Abelian subgroup $U(1)$ of $SU(2)$ in a covariant manner,  
\be{MAG}
{\cal L}_{AG}=\frac{F^a F^a}{2\alpha}-\cb^a M^{ab} c^b -g^2
\frac{\alpha}{2} (\cb^a\eps^{ab} c^b)^2\ ,
\ee
with 
\be{defs}
F^a= D_\m^{ab} W_\m^b=\partial_\m W_\m^a+ g A_\m\eps^{ab} W_\m^b\ {\rm
and}\ M^{ab}= D_\m^{ac} D_\m^{cb} + g^2\eps^{ac}\eps^{bd} W_\m^c
W_\m^d\ .
\ee
Like the corresponding Abelian LGT\cite{Sc99},
$L_{U(1)}=L_{\rm inv.}+{\cal L}_{\rm AG}$ is invariant under
$U(1)$-gauge transformations {\it and} an on-shell  
BRST symmetry $s$ and anti-BRST symmetry $\bar s$, whose action on the
fields is
\begin{displaymath}
\begin{array}{rclcrcl}
s A_\m&=&g\eps^{ab}c^aW_\m^b&\ \ \ \ & \bar sA_\m&=&g\eps^{ab}\cb^aW_\m^b\\
s W_\m^a&=& D_\m^{ab} c^b&& \bar s W_\m^a&=& D_\m^{ab} \cb^b\\
s c^a&=&0&&\bar s \cb^a&=&0\\
s \cb^a&=&F^a/\alpha&&\bar s c^a&=&-F^a/\alpha\ ,
\end{array}
\end{displaymath}
\vspace{-1.0cm}\be{brs}\ee
with an obvious extension to include matter fields. On the connections
$A_\m$ and $W_\m^a$ this BRST-variation effects an infinitesimal
transformation in the coset ${\cal G}/{\cal H}$ parameterized by the
two ghosts $c^a(x)$. Note that $s c^a=0$ here, because the coset
is not a group manifold.

The BRST algebra\equ{brs} closes on-shell 
on the set of $U(1)$-invariant functionals: on functionals that 
depend only on $W,A,c$ and the matter fields, $s^2$ for instance 
effects an infinitesimal U(1)-transformation with the parameter ${\frac g
2}\eps^{ab}c^a c^b$. The algebra \equ{brs} thus defines an
equivariant cohomology. It was derived from a more extensive 
nilpotent (off-shell) BRST-algebra on the lattice\cite{Sc99} by
integrating out some of the additional fields. As mentioned in the
introduction, the renormalizability and unitarity of this continuum
theory is guaranteed because it describes the critical limit of an
Abelian LGT that was proven
to have the same gauge invariant correlation functions as the original
$SU(2)$-LGT. Note that the physical sector comprises states created by
composite operators of $A,W$ and the matter fields in the equivariant
cohomology of $s$ (or $\bar s$). They are BRST closed,
$U(1)$-invariant and do not depend on the ghosts.  

For $\alpha>0$, \equ{MAG} could be considered a ``soft'' gauge fixing to
the Maximal Abelian Gauge (MAG). It differs from what one naively
obtains using a Faddeev-Popov procedure by a quartic ghost interaction
proportional to $\alpha$. \equ{MAG} also  
 does not implement the non-linear constraint $F^a=0$ exactly. 
However, setting $\alpha=0$  and perturbatively 
solving the constraint $F^a=0$ is {\it not} consistent and not  
the same as taking the limit $\alpha\rightarrow 0$. One could have 
inferred the highly singular nature of this limit from the fact that the 
4-ghost interaction diverges at one loop even when the photon- and 
vector boson propagators are transverse. A 4-ghost counterterm thus is
required even in the (formal) limit $\alpha\rightarrow 0$ and the
leading term in the  anomalous dimension of the gauge parameter in fact is $-3
g^2/(8\pi^2\alpha)$ in this limit\cite{Sc00}.
The physical reason for the singular behavior of the limit
$\alpha\rightarrow 0$  is inherently non-perturbative and nicely
exhibited by the lattice 
calculation\cite{Sc99}: without the quartic ghost interaction,
Gribov copies of a configuration conspire to give {\it vanishing} 
expectation values for all physical observables. No matter how small,
the quartic ghost interaction is required to have a
normalizable partition function and expectation values of physical
observables that are identical with those of the original
SU(2)-LGT. From a perturbative point of view, 
$\alpha\rightarrow 0$ at {\it finite} coupling $g^2$ corresponds to a strong 
coupling limit that is not accessible perturbatively\cite{Sc00}. 

${\cal L}_{\rm aGF}$ in\equ{L}  fixes the
remaining $U(1)$ gauge invariance and thus defines the perturbative 
series unambiguously. I will consider a conventional covariant
gauge-fixing of the form, 
\be{aGF}
L_{\rm aGF}=\delta [\bar\omega(\partial_\m A_\m-\frac{\xi}{2} b)] =
b\partial_\m A_\m -\frac{\xi}{2} b^2
+\bar\omega\bigtriangleup\omega\equiv\frac{1}{2\xi} (\partial_\m
A_\m)^2\ ,  
\ee
where the last equivalence is obtained by  decoupling the Abelian
ghosts $\omega$ and $\bar\omega$ and the Nakanishi-Lautrup field $b$.
$\delta$ is a BRST-symmetry defined on the fields as
\begin{displaymath}
\begin{array}{rclcrcl}
\delta A_\m&=&-\partial_\m \omega &\ \ \ \ & \delta W_\m^a&=& g\omega
\eps^{ab} W_\m^b\\ 
\delta c^a&=&g\omega \eps^{ab} c^b && \delta \cb^a&=&g\omega \eps^{ab}\cb^b\\
\delta \omega&=&0 &&&&\\
\delta\bar\omega&=&b && \delta b&=&0
\end{array}
\end{displaymath}
\vspace{-1.0cm}\be{delta}\ee
Trivially extending $s$ and $\bar s$ to the additional  fields
\be{ext}
s\omega=s\bar\omega=s b=\bar s\omega=\bar s\bar\omega=\bar s b=0\ ,
\ee
one can verify that $\delta$ is nil-potent and anticommutes with
$s$ and $\bar s$ 
\be{anti}
\delta^2=s \delta+\delta s=\bar s \delta+\delta \bar s=0
\ee
As far as algebraic renormalization is concerned, the action\equ{L}
thus is composed of a term in the cohomology of $\delta$ that is
invariant under the global symmetries $s$ and $\bar s$ given
in\equ{brs} and a $\delta$-exact term. The latter is not invariant
under $s$ nor $\bar s$. Since the global symmetries commute with
$\delta$, the situation is the same as in any gauge-fixing that 
breaks some of the global symmetries (or supersymmetries) of the
theory.  There is a well-defined procedure to handle this
case\cite{Pi95} in algebraic
renormalization. From a more heuristic point of view, we already know
that the $s$ and $\bar s$ symmetries of the theory are not anomalous
from the lattice regularization\cite{Sc99} of this model. 
I will therefore not give the algebraic proof here. 

What has been gained compared to conventional covariant gauge fixing?
Since the present gauge fixing in a sense is ``hierarchical'', we are 
able to single out the maximal Abelian subgroup. As 
emphasized before, the $s$ and $\bar s$ symmetries can be implemented
on the lattice and the resulting Abelian LGT shown to be physically
equivalent to one with an SU(2) structure group. It
may eventually be possible to construct the dual of this Abelian
LGT. In addition, the theory described by\equ{L} does not suffer from a
generic Gribov problem due to zero modes of the ghosts. Since the
global $s$ and $\bar s$ symmetries are preserved by the lattice
regularization, \equ{L} probably descibes the critical limit
of a LGT better than any other set of covariant gauges. Finally,
because the Abelian $\omega$-ghost and
$\bar\omega$-antighost as well as the Nakanishi-Lautrup field $b$
decouple, we effectively end up with a local and covariant gauge-fixed
theory with fewer ghosts. This reduction in the number of fields is at the
expense of a quartic ghost interaction that gives the ghosts 
some interesting dynamics of their own.  

\section{The Dynamically Broken SL(2,R) Symmetry}
The Lagrangian \equ{L} also exhibits a global bosonic
SL(2,R) symmetry that is generated by  
\be{SL2}
\Pi^+=\int c^a(x)\frac{\delta}{\delta \cb^a(x)}\ \ ,\ \  
\Pi^-=\int \cb^a(x)\frac{\delta}{\delta c^a(x)}\ ,
\ee
and the ghost number $\Pi=[\Pi^+,\Pi^-]$. This SL(2,R) symmetry is 
preserved by the regularization (for instance dimensional) and thus is
not anomalous. The conserved currents corresponding to $\Pi^\pm$ 
are U(1)-invariant and BRST, respectively anti-BRST exact,
\be{currents}
j^+_\m=c^a D^{ab}_\m c^b=s c^aW^a_\m\ ,\ \ j_\m^-=\cb^a
D^{ab}_\m\cb^b=\bar s \cb^aW_\m^a\ . 
\ee
I will argue\cite{Sc00} that the global $SL(2,R)$ symmetry of the
theory is spontaneously broken to the Abelian subgroup
generated  by the ghost number $\Pi$. 
An order parameter for the spontaneous breakdown of the
SL(2,R) symmetry thus is
\be{opar}
\bra\cb^a\eps^{ab}c^b\ket=\half\bra\Pi^-
(c^a\eps^{ab}c^b)\ket=-\half\bra\Pi^+ (\cb^a\eps^{ab} \cb^b)\ket\ . 
\ee

Because the
currents\equ{currents} are (anti)-BRST exact, a
spontaneously broken SL(2,R) symmetry is accompanied by a BRST-quartet
of massless Goldstone states with ghost numbers $2,1,-1$ and $-2$. They are
U(1)-invariant $c-c$, $c-W$, $\cb-W$ and $\cb-\cb$ bound states. 
It is important to note in this context that BRST quartets do not
contribute to physical quantities\footnote{This is analogous to the
decoupling of the Goldstone quartets of the weak interaction  in
renormalizable $R_\xi$ gauges\cite{Ku78}.}. The spontaneous breaking
of the $SL(2,R)$ symmetry  
in a sense is similar to a dynamical Higgs mechanism in the adjoint. 
The vector bosons $W$ aquire a mass (see below) but in 
contrast to a conventional Higgs mechanism in the adjoint, this mass is 
not a free parameter of the theory, but can be determined in terms of 
$\LQCD$.     

To see that ghost condensation will almost invariably occur at weak
coupling in the model described by\equ{L} it is illustrative to compare 
with the BCS-theory\cite{Schrieffer99} of superconductivity. 
In BCS-theory, an at low momentum transfers attractive  
4-fermion interaction leads to the condensation of certain fermion
pairs and the formation of a gap in the quasi-particle 
spectrum near the Fermi surface. 
An analogous phenomenon occurs here for ghost and anti-ghost 
modes corresponding to small eigenvalues of the FP-Operator $M^{ab}$
defined by  \equ{defs}: for zero modes, the bilinear term in\equ{MAG}
vanishes and the quartic ghost interaction selects the channel in
which condensation occurs. The quartic ghost interaction in\equ{MAG}
is attractive  when the color of the ghost and anti-ghost are
opposite: it thus leads to the formation of a
$\vev{\cb^a\eps^{ab} c^b}$ condensate at arbitrarily weak coupling
$\alpha g^2$ by the BCS-mechanism. [The two spin states of a fermion
here has an analog in the two color orientations 
of the ghosts. We choose ghost number to be conserved and observe 
$\cb-c$, rather than $c-c$ or $\cb-\cb$  condensation, as would be the
case if we chose $\Pi^-$ or $\Pi^+$ as unbroken generators.] 
The analogy with BCS-theory is particularly
appealing because the operator $M^{ab}$ has small eigenvalues 
whenever the gauge field configuration is in the vicinity of a Gribov
horizon.  That the ground state may be dominated by such
configurations was previously suggested in an 
attempt to restrict the functional integral to the fundamental modular
region\cite{Zw92} of Landau gauge. In the present context, gauge field
configurations  with non-Abelian monopoles are on the Gribov horizon,
since the  {\it failure} of the gauge fixing condition to an Abelian
subgroup is necessary  for the presence of monopoles. 

Thus, if monopoles are relevant in describing the ground state of
the theory, it is not inconceivable that the ghosts will condense. The
converse is not necessarily true because $M^{ab}$ can  
have arbitrarily small eigenvalues at field configurations with 
vanishing monopole number. We will see below that  the ghosts already
condense in the vicinity of the trivial gauge field configuration. 
The analogy with BCS-theory suggests that they condense 
for {\it any} value of the quartic coupling $\alpha g^2$, with a gap
that depends exponentially on $1/(\alpha
g^2)$. 

To perturbatively investigate the consequences of
$\bra\cb^a\eps^{ab}c^b\ket\neq 0$, the quartic ghost
interaction in\equ{MAG} is linearized by introducing an auxiliary scalar
field $\rho(x)$ of canonical dimension two. Adding the quadratic term 
\be{Stratanovic}
{\cal L}_{\rm aux}=\frac{1}{2 g^2} (\rho -g^2\lambda \cb^a\eps^{ab} c^b)^2
\ee
to the Lagrangian of\equ{L}, the tree level quartic ghost interaction
 vanishes at $\lambda^2=\alpha $
and is then formally of $O(g^4)$, proportional to the difference 
$Z^2_\lambda-Z_\alpha$ of the renormalization constants of the two 
couplings\footnote{The discrete symmetry $ c^a\rightarrow \cb^a,\
\cb^a\rightarrow -c^a,\ \rho\rightarrow -\rho$ relating $s$ and $\bar
s$ also ensures that $\rho$ only mixes with $\cb^a\eps^{ab}c^b$.}.

We will see that the perturbative expansion
about a {\it non-trivial} solution  $\langle\rho\rangle=v\neq 0$ to the 
gap equation
\be{gap}
\frac{v}{g^2}=\sqrt{\alpha}\left.\bra c^a(x)\eps^{ab}{\bar
c}^b(x)\ket \right|_{<\rho>=v}\ ,
\ee
is stable and much better behaved in the
infrared. 

Defining the quantum part $\sigma(x)$ of the auxiliary scalar $\rho$ by
\be{decomp}
\rho(x)=v+\sigma(x) \ \ {\rm with}\ \ \bra\sigma\ket=0\ ,
\ee
the momentum representation of the Euclidean ghost propagator at 
tree level becomes  
\be{ghost}
\bra c^a \cb^b\ket_p=\frac{p^2\delta^{ab}+\sqrt{\alpha}v\eps^{ab}}{p^4+\alpha v^2}=\int_0^\infty \!\! d\omega\,
e^{-\omega p^2}[\delta^{ab}\cos(\omega v\sqrt{\alpha})+\eps^{ab}\sin(\omega
v \sqrt{\alpha})]\ .
\ee
Feynman's parameterization of this propagator leads to an 
evaluation of loop integrals using dimensional regularization  that is
only slightly more complicated than usual. 

Using\equ{ghost} in\equ{gap} the gap
equation to one-loop in $d=4-2\eps$ dimensions is, 
\ba{gapt}
\frac{v}{\hat g^2}&=&2\mu^4\sqrt{\alpha}\int_0^\infty
\frac{d\omega}{(4\pi\mu^2\omega)^{2-\eps}} \sin(\omega\sqrt{
v^2\alpha})\nonumber\\
&=&\frac{\alpha v}{8\pi^2}\left[\frac{1}{\eps} -\ln\frac{\pi y^2
T^2}{\mu^2 e^{1-\gamma_E}} +O(\eps)\right]\ .
\ea
Here $\gamma_E$ is Euler's constant. Including the counterterm
\be{counter}
\frac{v }{\hat g^2} \rightarrow \frac{v }{\hat g^2} Z_v^2 Z_g^{-2} =  
\frac{v }{\hat g^2} +\frac{\alpha v}{8\pi^2\eps}+O(\hat g^2)
\ee
on the left-hand side of\equ{gapt} cancels the $1/\eps$ divergence of
the right-hand side of\equ{gapt}. The renormalized  (non-trivial)
solution to the gap equation in four 
dimensions thus is,
\be{sol}
\alpha v^2=e^2 \Lambda^4,\ {\rm with}\ \ \Lambda^2(\alpha, g,\mu)=4\pi\mu^2 e^{-\gamma_E -\frac{8\pi^2}{\alpha
g^2}}\ .
\ee
Note the exponential dependence of the gap $v$ on the quartic coupling
$\alpha g^2$ expected from BCS-theory. One can show that the
solution\equ{sol} corresponds to the global minimum of the effective
potential by either directly computing the (one-loop renormalized) effective
potential, 
\be{veff}
V(v,\m, g,\alpha)=\frac{\alpha v^2}{32\pi^2}\ln\frac{\alpha
v^2}{e^3 \Lambda^4} +O(g^2)\ ,
\ee
or by integrating\equ{gapt} and noting that
$V(\Lambda,0)=0$. Consistency requires that the 1PI two-point function of
the scalar $\sigma$ is positive definite for all Euclidean momenta
when\equ{gap} is satisfied\cite{Sc00}.  
From\equ{counter} one obtains for the anomalous dimension of $v$ to one loop
\be{anom}
\gamma_v=-\frac{d \ln Z_v}{d \ln
\mu}=\frac{g^2}{16\pi^2}(2\alpha-\beta_0) +O(g^4)
\ee
where $\beta_0$ is the lowest order coefficient of the
$\beta$-function. At $\alpha=\beta_0/2$, the anomalous
dimension of $v$ is of order $g^4$ and corrections to the
asymptotic $v(g\rightarrow 0)$ solution in this particular {\it
critical} gauge therefore are {\it analytic} in $g^2$ and may be
computed order by order in perturbation theory. In this {\it critical} gauge 
we thus have that the scale $\Lambda$ describing the minimum of the
effective potential $V(v,\mu,g,\alpha)$ is analytically related to $\LQCD$:
\be{crit}
\Lambda(\alpha=\beta_0/2,g,\mu)=\LQCD (1+O(g^2))
\ee
In other gauges $v\neq 0$ at weak coupling is either much larger
than $\LQCD$ (for $\alpha\gg\beta_0/2$), or much smaller (for
$\alpha\ll \beta_0/2$). To leading order in the loop expansion, the anomalous
dimension\equ{anom} does not vanish in these cases and higher order
loop corrections to $v\neq 0$ are of comparable magnitude. 
[As noted before, the extreme limit $\alpha\rightarrow 0$
in which the non-trivial solution\equ{sol} becomes degenerate with the trivial
one, in particular corresponds to a strong coupling problem.] In the
critical gauge $\alpha=\beta_0/2$, the 
perturbative 1-loop calculation of $v\neq 0$ is self-consistent in
the sense that all higher order corrections to the expectation value 
are of order $g^2$ because the anomalous
dimension\equ{anom} is of order $g^4$ at this point. I wish to
emphasize that this does {\it not} imply that physical effects
associated with ghost condensation in this gauge are themselves 
gauge dependent. It only implies that a non-trivial
solution to the gap equation is {\it perturbatively} consistent at 
$\alpha=\beta_0/2$. [That some gauges are better
suited than others for a non-perturbative evaluation of gauge invariant 
quantities is well known from QED: the gauge invariant
hydrogen spectrum is qualitatively best obtained in Coulomb
gauge. Below we relate $\alpha v^2$ to the vacuum expectation value of
the trace of the energy-momentum tensor.]
    
\section{The Vector Boson Mass}
Recent lattice simulations\cite{Am99} indicate that the
$W$-bosons are massive in maximal Abelian gauges. At least
qualitatively, this may be explained by the mechanism discussed here. Although
the tree level 
contribution to $m_W^2$ vanishes by Bose symmetry, 
ghost condensation induces\cite{Sc00} a {\it finite} mass
$m^2_W=g^2\sqrt{\alpha v^2}/(16\pi)$ at one loop as shown in Fig.~1. 
\vskip 1truecm
\vbox{
\be{Wmass}
\hskip 4truecm {=\frac{g^2\sqrt{v^2\alpha}}{16\pi}\delta_\mn \delta^{ab}\ .}
\ee 
\vskip-2.0truecm
\hskip .5truecm\psfig{figure=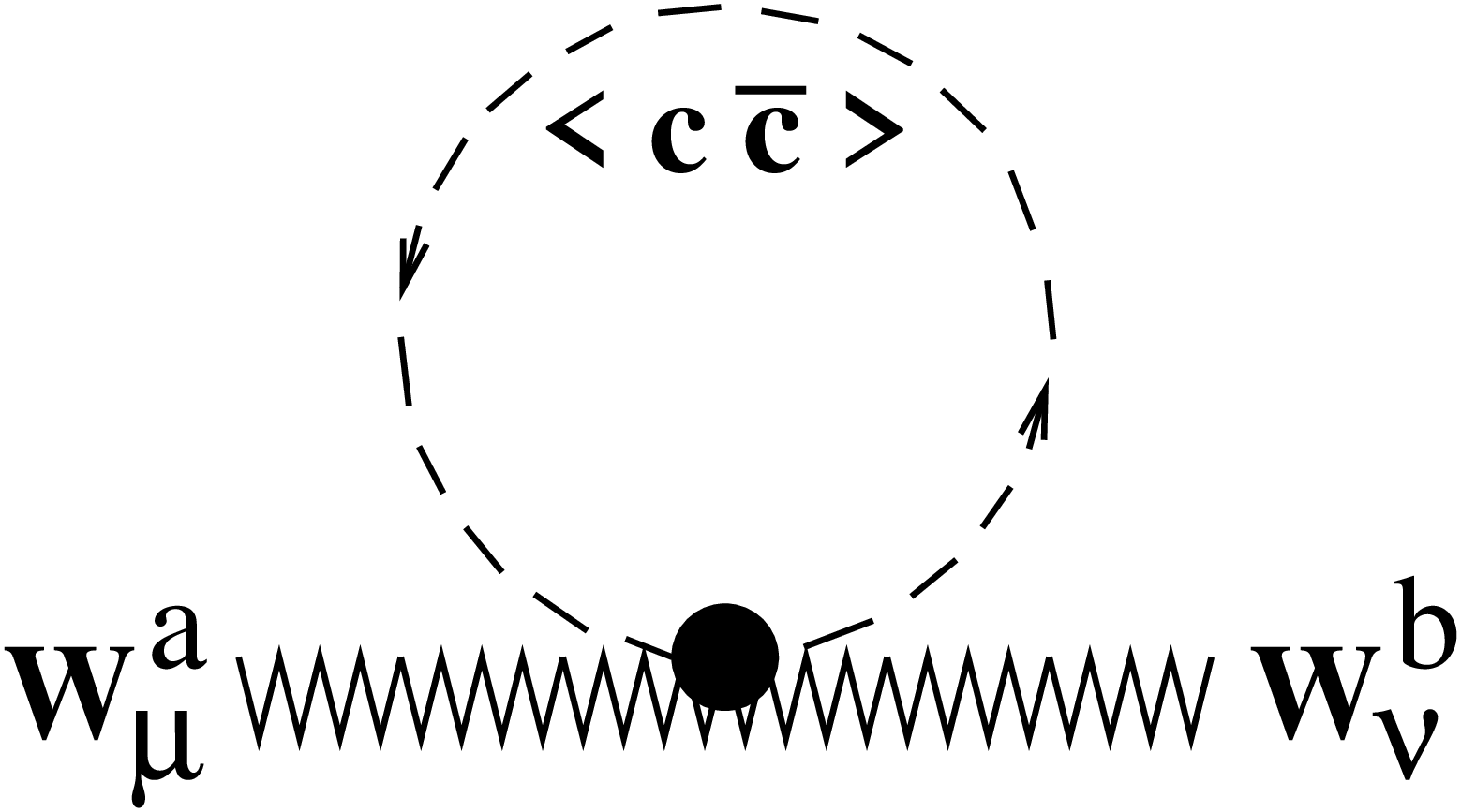,height=2.0truecm}
\\*
{\small\noindent Fig.~1. The finite one-loop contribution to the
$W$ mass.}}
\vskip 5pt
Technically, the one-loop contribution is finite because the integral 
in\equ{Wmass} involves only the $\delta^{ab}$-part
of the ghost propagator\equ{ghost}.
Since $p^2/(p^4+\alpha v^2)=-\alpha v^2/(p^2(p^4+\alpha v^2))+1/p^2$, the 
$v$-dependence of the loop integral is IR- {\it and}
UV-finite. The quadratic UV-divergence of the $1/p^2$ 
subtraction at $v=0$ is canceled by the other,
$v$-independent, quadratically divergent one-loop contributions -- (in
dimensional regularization this scale-invariant integral 
vanishes by itself). $m_W^2$ furthermore is {\it positive}
due to the overall minus sign of the {\it ghost} loop. The sign of $m_W^2$ is
crucial. It is a further indication that the model is {\it
stable} and (as far as the loop expansion is concerned)
does not develop tachyonic poles at Euclidean $p^2$ for $v\neq 0$. 
Conceptually, the local mass term proportional to
$\delta_\mn\delta^{ab}$ is finite due to the
BRST symmetry\equ{brs}, which excludes a mass counter-term. 
The latter  argument implies that contributions to  $m^2_W$ are finite to all
orders of the loop expansion. 

Although this ``mass''-term regulates the IR-behavior of the
$W$-propagator perturbatively, the 1-loop calculation above 
should quantitatively describe the behavior of the $W$-propagator 
at high momenta, where $g^2(p^2)$ is a small parameter and this
calculation is consistent. Ghost condensation thus should lead to 
a leading power correction $\propto v/p^2$ to the $W$-propagator at high
momenta. The consistency of this behavior with the Operator Product
Expansion (OPE) is examined below. 

\section{Discussion of Physical Consequences}
We have seen that ghost condensation in covariant Abelian gauges is
associated with the spontaneous breaking of a global SL(2,R) symmetry
whose diagonal generator is the ghost number. The currents of the
broken symmetries are (anti-)BRST exact and the Goldstone states 
form a BRST-quartet that decouples from physical observables. What are
observable consequences? We gain some insight by 
computing the contribution to the expectation value of the  
trace of the energy momentum tensor. From the effective 
potential\equ{veff} one obtains: 
\be{trace}
\bra\theta^\m_\m\ket=-\frac{\alpha v^2}{8\pi^2}=-\frac{e^2}{8\pi^2}\Lambda^4 , 
\ee
At the minimum of the effective potential. 
$v\neq 0$ thus lowers the vacuum energy density and ghost
condensation may be interpreted as a low-energy manifestation of the 
trace anomaly.  

Let me also comment on the consistency of the approach from the point 
of view of the Operator Product Expansion (OPE). In generic covariant 
gauges, the OPE implies that power corrections to physical correlators 
(but also to Green functions\cite{Ah92}) are at
least suppressed by an order $M^4/p^4$ relative to the perturbative
behavior at high momenta (in the absence of quarks). This is simply
because the operator of lowest dimension in the BRST-cohomology
of these gauges has canonical dimension four. Consistency of the
present approach requires that one explain
\begin{itemize}
\item why the ground state in the present case {\it can} support a
vacuum expectation value $v\propto\vev{\cb^a\eps^{ab} c^b}\neq 0$ of
canonical dimension two and simultaneously,
\item why power corrections of order $v/p^2$ are absent in {\it
physical} correlation functions
\end{itemize}
To address the 
first question, we construct a $U(1)$-invariant local operator of
dimension two whose vacuum expectation value manifestly is invariant
 under the BRST-algebra of\equ{brs}. Note that
\ba{op}
s [\alpha \cb^a c^a +\half W_\m^a W_\m^a] &=&\partial_\m (W_\m^a
c^a)\nonumber\\
\bar s [\alpha \cb^a c^a +\half W_\m^a W_\m^a] &=&\partial_\m (W_\m^a
\cb^a)\ .
\ea
$O_2:=[\cb^a c^a +\frac{1}{2\alpha} W_\m^a W_\m^a]$ 
is {\it not} in the equivariant cohomology of the on-shell 
BRST-algebra\equ{brs}, because $s^2\sim 0$ only on U(1)-invariant
operators that do {\it not} depend on $\bar c$ whereas $O_2$
does. Nevertheless, the zero-momentum component of $O_2$ is invariant
under $s$, $\bar s$ and $\delta$. $\vev{O_2}\neq 0$ therefore is
an invariant statement as far as the algebra of symmetries is
concerned. We have seen that $\vev{O_2}=\frac{\sqrt{\alpha
v^2}}{16\pi}\neq 0$ in fact appears to be the dynamically favored
possibility. The existence of an operator of dimension two whose
zero-momentum component is invariant explains the leading power
corrections we find for the $W$- $A$- and ghost- propagators from the point
of view of the OPE. 

\hskip 1.truecm\psfig{figure=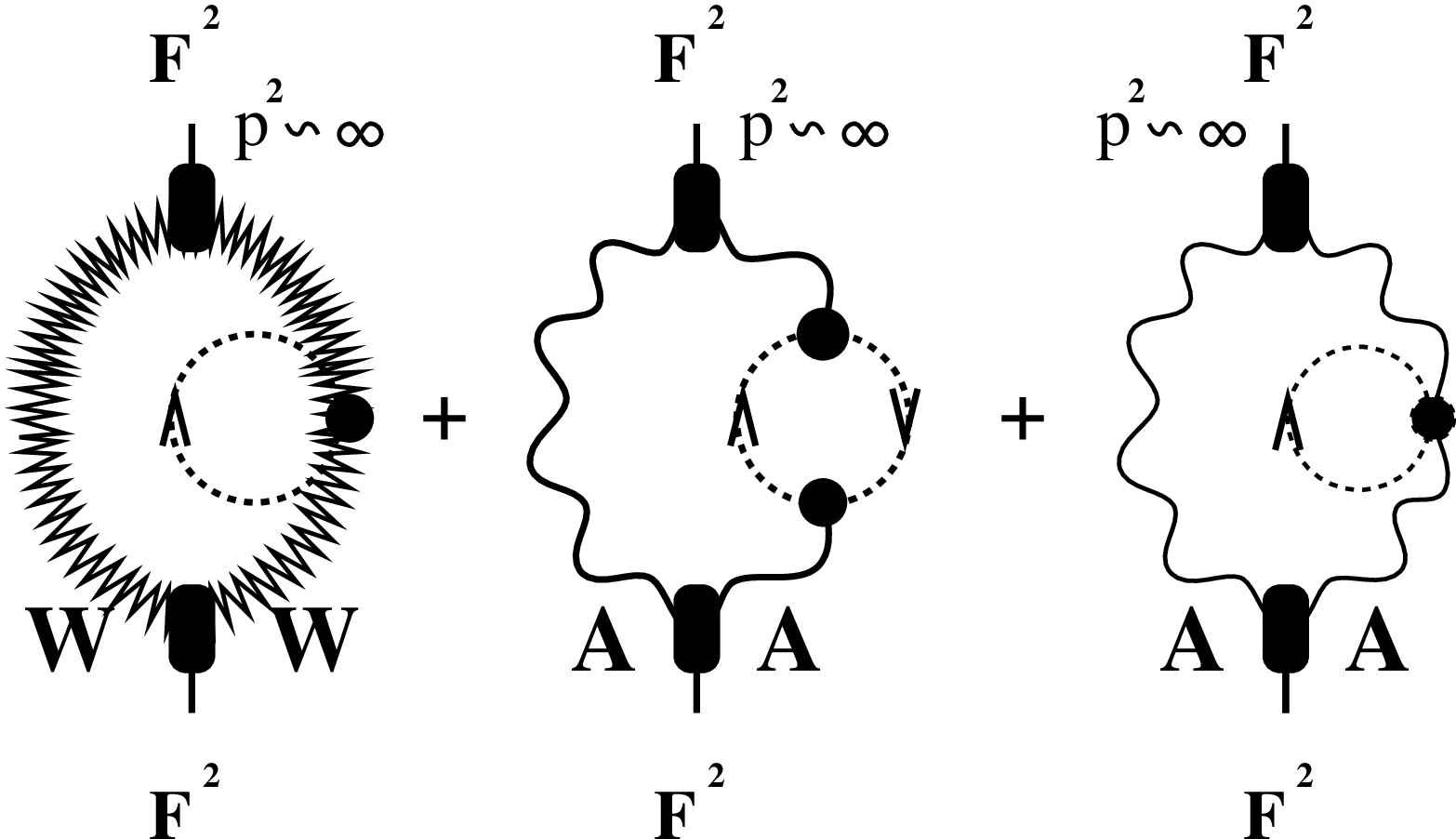,height=4.0truecm}
\\*
{\small\noindent Fig.~2. Schematic representation of leading  
contributions that give rise to power corrections at
large momenta  in the gauge invariant correlators \equ{corr}.
 For reasons given in the main text, the power leading correction
$\propto g^2 v$ of the propagators cancels in this gauge invariant 
combination.}
\vskip 5pt

On the other hand, since $O_2$ is {\it not} invariant under {\it global}
$SU(2)$ transformations, we do not expect $\vev{O_2}$ to appear in the
OPE of gauge invariant correlators. The leading power correction
$\propto v$ in the propagators therefore should 
cancel in gauge invariant correlation functions such as
\be{corr}
\vev{(G_\mn G_{\rho\sigma}+G^a_\mn G^a_{\rho\sigma})
(G_{\alpha\beta} G_{\gamma\delta}+G^b_{\alpha\beta}
G^b_{\gamma\delta})}_{p\sim\infty}
\ee
To leading order in the loop expansion this can be verified
explicitly. To this order,  the three 
diagrams of Fig.~2 with two (transverse) photons and two
(transverse) vector bosons as intermediate states lead to 
power corrections. In the limit
$p^2\sim\infty$ at least one of the photons, respectively vector
bosons in the loop integrals has momentum much larger than $v$. The
leading power correction $\propto v$ in\equ{corr} from the
vector bosons and the photon thus cancel if and only if the photon
 polarization,
\be{pol}
\Gamma^{AA}_\mn(p,v)=(\delta_\mn p^2 -p_\m p_\n) \Pi^{AA}(p^2,v)
\ee
for $p^2\gg v$ has the asymptotic power expansion
\be{polas}
\Pi^{AA}(p^2\gg v, v)\sim 1 -2 \frac{m^2_W}{p^2}+O(g^2\ln p^2 ,v^2/p^4)\ .
\ee 
Here $m^2_W=g^2 \frac{\sqrt{\alpha v^2}}{16\pi}$ is the $W$-boson
mass\equ{Wmass}. Evaluating the one ghost loop contributions to
the photon polarization at high external momentum  one indeed
verifies\equ{polas}. Note that the factor $-2$ in\equ{polas} is
essential for the cancellation of the leading power correction 
in\equ{corr}, since twice as many $W$'s as photons contribute.
Power corrections of order $\alpha v^2/p^4$ in the asymptotic
expansion of\equ{corr} do {\it not} cancel and are related to
$\vev{\theta^\m_\m}$ by \equ{trace}.

We have here proposed a mechanism by which the $W$-bosons of an SU(2)
gauge theory in Abelian gauges essentially become massive while the 
leading power corrections to gauge-invariant correlation functions 
nevertheless are of order $\LQCD^4/p^4$
only. Although numerical lattice simulations\cite{Am99} show similar
effects for the off-diagonal gluons, the numerical  gauge fixing to
MAG is not described by a local effective 
action and the results therefore cannot be 
directly compared with the ones presented here. One
unfortunately cannot even extract the anomalous dimension of 
$m_W^2$ from the present numerical studies due to their rather narrow range of
couplings. Although lattice simulations presently do not unambiguously 
confirm the mechanism of mass generation by ghost
condensation I discussed, the simplicity and inherent consistency
of this approach may warrant further study.


\section*{References}
\begin{thebibliography}{99}

\bibitem{Fo95}  P.de~Forcrand and J.E.Hetrick, {\it
Nucl. Phys. Proc. Suppl.} {\bf 42} (1995) 861.

\bibitem{Pa99}  A.Patrascioiu and E.G.Seiler, {\it Difference between
abelian and nonabelian models: Fact and Fancy}, MPI preprint 1991,
math-ph/9903038.

\bibitem{Sc99} M. Schaden, {\it Phys. Rev.} {\bf D59} (1999) 014508.

\bibitem{Se82} E.G.Seiler, {\it Gauge theories as a problem of
constructive field theory and statistical mechanics} (Springer, New
York 1982).

\bibitem{Sc00} M.Schaden, {\it Mass Generation in Continuum SU(2)
Gauge Theory in Covariant Abelian Gauges}, hep-th/9909011.

\bibitem{Pi95} O.Piguet and S.P.Sorella, {\it Algebraic
renormalization: perturbative renormalization, symmetries and
anomalies} (Springer, New York 1995).

\bibitem{Ku78} T. Kugo, I. Ojima, {\it Prog. Theor. Phys.} {\bf 60}
(1978) 1869; {\it ibid.}{\bf 61} (1979) 294.

\bibitem{Schrieffer99} J.R. Schrieffer, {\it Theory of
Superconductivity} (Oxford, Perseus 1999). 

\bibitem{Zw92} D.Zwanziger, {\it Nucl. Phys.} {\bf B378} (1992) 525.

\bibitem{Am99} K.Amemiya and H.Suganuma,{\it Phys. Rev. } {\bf D60}
(1999) 114509. 

\bibitem{Ah92} J.Ahlbach, M.Lavelle, M.Schaden, A.Streibl, {\it
Phys. Lett.} {\bf B275} (1992) 124.

\end{thebibliography}
\end{document}